\documentstyle[pra,aps,twocolumn,epsf]{revtex}

\begin{document}
\draft
%\preprint{cond-mat}
\twocolumn[\hsize\textwidth\columnwidth\hsize\csname @twocolumnfalse\endcsname

\title{
Novel mechanism of photoinduced reversible phase transitions
in molecule-based magnets
}

\author{Tohru Kawamoto, Yoshihiro Asai, and Shuji Abe}
\address{
Electrotechnical Laboratory, Agency of Industrial Science and Technology (AIST), 1-1-4 Umezono, Tsukuba 305-8568, Japan.}
\date{\today}
\maketitle

%
% abstract
%

\begin{abstract}
A novel microscopic mechanism of bi-directional structural
changes is proposed for the photo-induced magnetic phase transition
in Co-Fe Prussian blue analogues
on the basis of ab initio quantum chemical cluster calculations.
It is shown that the local potential energies of various spin states of
Co are sensitive to the number of nearest neighbor Fe vacancies.
As a result, the forward and backward structural changes are
most readily initiated by excitation of different local regions
by different photons.
This mechanism suggests an effective strategy to realize
photoinduced reversible phase transitions in a general system
consisting of two local components.
\end{abstract}
\vskip2pc]
%\vskip2pc

\narrowtext

%::::::::::::::::::::::::::::::::::::::::::::::::::::::::::::::::::::::::::::::
%                              introduction
%::::::::::::::::::::::::::::::::::::::::::::::::::::::::::::::::::::::::::::::

Repeatable switching-on and -off of magnetization by external stimuli
such as light is one of the most fascinating phenomena with potential
applications in next generation's information storage and processing.
A bi-directional photo-induced magnetization
was first discovered in a cobalt-iron Prussian blue analogue,
K$_{0.4}$Co$_{1.3}$Fe(CN)$_6\cdot$ 5H$_2$O\cite{Sato97JES,Sato96Science,Verdaguer96Science}.
Illumination of visible light (500 - 700 nm) at low temperature induces a
bulk magnetization (presumably, ferrimagnetism), which can be eliminated by
illumination of near-IR light ($\sim$1300 nm).
In spite of various experimental and theoretical 
efforts\cite{Sato97JES,Sato96Science,Goujon00EPJB,Pejakovic00PRL,Yoshizawa98JPCB,Kawamoto99PRB,Nishino98PRB}, 
the microscopic mechanism of the reversible magnetization is still not clear.
In this Letter we report {\em ab initio} quantum chemical cluster calculations
for the Co-Fe Prussian blue analogue,
unveiling the microscopic mechanism of the bi-directional photo-induced
local structural changes that trigger the phase transitions.

\begin{figure}
\begin{center}
  \leavevmode
  \epsfxsize=80mm
  \epsfbox{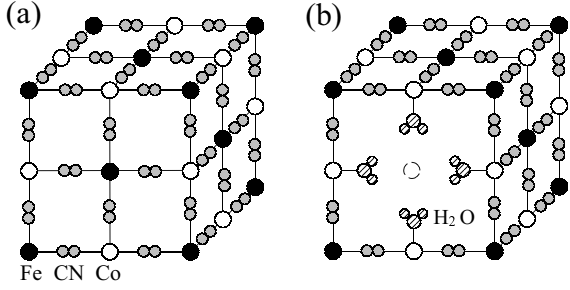}
\end{center}
\caption{
Crystal structure of K$_{1-2x}$Co$_{1+x}$Fe(CN)$_6$: (a) $x=0$; (b) $x\neq 0$.
Black, white, and gray circles denote Fe, Co and CN, respectively.
In (b), there is an vacancy at an iron site
surrounded by six water molecules substituting cyano anions.
}
\label{fig:crystal}
\end{figure}

The crystal of a Prussian blue analogue K$_{1-2x}$Co$_{1+x}$Fe(CN)$_6$ 
is composed of two metallic sites 
located on vertices of the cubic lattice and each
surrounded by six cyano moieties, as shown in Fig.~\ref{fig:crystal}.
The $d$-orbitals of transition metals split into $t_{\rm 2g}$ and $e_{\rm g}$
orbitals by the ligand field.
In the case of x$\neq$0,
there are vacancy sites with replacement of CN
by H$_2$O as shown in Fig.~\ref{fig:crystal}(b).
Various fascinating phenomena including
a room temperature magnet\cite{Ferlay95Nature},
electrochemically tunable magnets\cite{Sato96Science2},
transparent and colored magnetic thin films\cite{Ohkoshi98JACS},
and photo-induced magnetic dipole inversion\cite{Ohkoshi99JACS} 
have been observed
in such non-stoichiometric compounds. 
We will show that this non-stoichiometric aspect is essential for the
reversible photo-induced magnetization. 

\begin{figure}
\begin{center}
  \leavevmode
  \epsfxsize=80mm
  \epsfbox{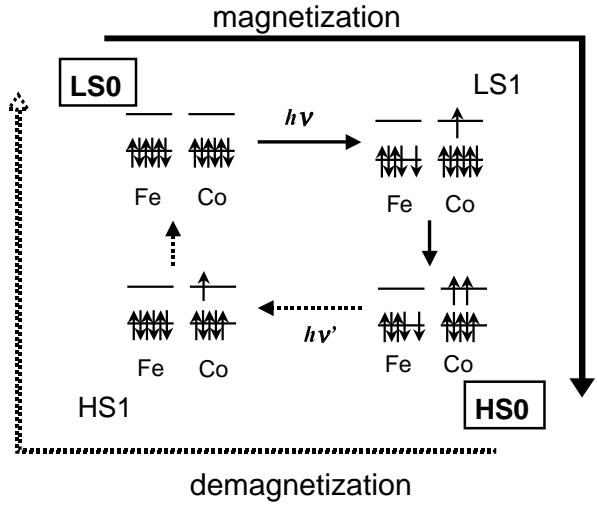}
\end{center}
\caption{
Schematic local spin configurations of a Co-Fe pair.
LS0, LS1, HS0, and HS1 represent the non-magnetic ground state, the
paramagnetic excited state, the magnetic meta-stable state, and the magnetic
excited state, respectively.
The magnetization LS0$\rightarrow$HS0 occurs through charge transfer
excitation to LS1 by photon $h\nu$ and subsequent intersystem crossing
(I.C.), whereas
the demagnetization HS0$\rightarrow$LS0 occurs similarly via HS1.
}
\label{fig:cycle}
\end{figure}

The low spin (LS) configuration of the ground nonmagnetic state and the high
spin (HS) configuration of the meta-stable magnetic state of
K$_{0.4}$Co$_{1.3}$Fe(CN)$_6\cdot$ 5H$_2$O are most likely
Co$^{\rm III}(d\varepsilon^6,S=0)
$Fe$^{\rm II}(d\varepsilon^6,S=0)$ and
Co$^{\rm II}(d\varepsilon^5d\gamma^2,S=3/2)
$Fe$^{\rm III }(d\varepsilon^5,S=1/2)$,
respectively\cite{Sato97JES,Sato96Science}.
These are depicted in Fig.\ref{fig:cycle} as LS0 and HS0 states.
Figure \ref{fig:cycle} schematically represents the most plausible
elementary processes in the cycle of the photo-induced structural change.
The LH0 and HS0 states are converted to the intermediate states LS1 and HS1,
respectively, by photo-induced charge transfer (CT) between iron and cobalt
atoms, and then to the final states HS0 and LS0 by intersystem crossing due
to spin-orbit coupling at cobalt sites.
The major structural difference between the LS and HS states
lies in the Co-N bond length
$d$(Co-N), which is longer by about 0.2\AA~ in HS
than in LS\cite{Yokoyama98PRB}.
This implies that the transition between LS and HS states requires 
a volume change, which may be a major origin of a large energy barrier 
between them, as in the case of a spin-crossover complex\cite{Decurtins84CPL,Ogawa00PRL}.

%::::::::::::::::::::::::::::::::::::::::::::::::::::::::::::::::::::::::::::::
%                       method and procedure of calculations
%::::::::::::::::::::::::::::::::::::::::::::::::::::::::::::::::::::::::::::::

Based on these observations, we adopt a cluster model consisting of 
a cobalt ion and surrounding cyano ligands with a variable bond length 
between Co and N. 
Previously we carried out first-principles band structure 
calculations\cite{Kawamoto99PRB}.
It was found that the relevant d-bands have quite narrow widths compared
with the level spacings, implying that the d-electrons are fairly localized
with weak charge transfer character in this material.
This allows us to use the cluster approach in the present paper.
We have checked that cluster calculations produce results that are
consistent with band structure calculations\cite{Kawamoto99PRB}.

To take into account the partial substitution of ligands by water molecules, 
we consider the following three clusters with various numbers $N_{\rm W}$ of 
H$_2$O substitutions:
(a) Co(NC)$_6$ ($N_{\rm W}$=0),
(b) Co(NC)$_5\cdot$H$_2$O ($N_{\rm W}$=1),
and
(c) Co(NC)$_4\cdot2$H$_2$O ($N_{\rm W}$=2).
Bond lengths other than Co-N are fixed at observed values, e.g., 
$d$(C-N)=1.12\AA\cite{Yokoyama98PRB}. 
The potential energies of the four relevant states shown in Fig.\ref{fig:cycle} 
 are calculated for these clusters
by using a quantum chemical ab-initio calculation program 
with the MIDI basis function set\cite{Schmidt93JCompPhys,Huzinaga84book}.
The multi-configuration self-consistent field (MCSCF) method is used with 
an active space limited to five orbitals mainly composed 
of Co-$d$ atomic orbitals.
The spin orbit coupling was not included 
to obtain the energy curve for each spin state.

As for the treatment of surrounding atoms outside the cluster,   
we use a lattice of atomic point charges evaluated from 
our previous band calculation \cite{Kawamoto99PRB}.
The Fe-C bond length is fixed at the observed value 1.91\AA\cite{Yokoyama98PRB}. 
The environment lattice contains one or two iron vacancies 
situated just outside the water molecules of the clusters (b) and (c).
The energies of the four states are calculated for each value of $d$(Co-N) 
common to all the Co-N bonds in the cluster as well as in the 
environmental lattice.
This choice diminishes the interfacial strain energy 
due to local volume change,  
being a suitable way to extract the tendency of the local cluster.
It should be noted that the real relaxation dynamics  
is determined not only by the cluster potentials but also by (mainly elastic) 
interactions between clusters.

Since each Fe has full six coordination of CN with the fixed Fe-C distance, we
assume that the energy difference between the two states
Fe$^{\rm II}(d\varepsilon^6,S=0)$ and
Fe$^{\rm III}(d\varepsilon^5,S=1/2)$ is a constant $\Delta$ irrespective of
Co states.
Then the energy difference between the LS0 and HS0 states, for example, of a
Co-Fe pair is written in an additive form:
$\Delta E_{{\rm LS0}-{\rm HS0}}$=$E$[Co$^{\rm III} (d\varepsilon^{6},S=0)] 
- E$[Co$^{\rm II}  (d\varepsilon^{5} d\gamma^{2},S=3/2)]  - \Delta$.
We have checked the appropriateness of this approximation by 
a supplementary calculation for a two-metal cluster
Co(NC)$_5$-(NC)-Fe(CN)$_5$. 
In the following we use $\Delta$=0.4eV.
This choice ensures that the ground state is in LS state at x=0.3\cite{Sato97JES} 
and in HS state at x=0.5\cite{Juszczyk94JPhys} as observed experimentally.

%::::::::::::::::::::::::::::::::::::::::::::::::::::::::::::::::::::::::::::::
%                                    results
%::::::::::::::::::::::::::::::::::::::::::::::::::::::::::::::::::::::::::::::

\begin{figure}[h]
\begin{center}
  \leavevmode  \noindent
  \epsfxsize=70mm
  \epsfbox{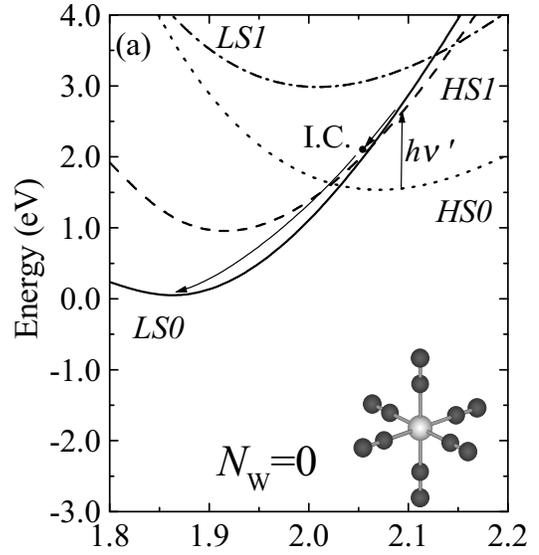}
\hspace{6mm}
  \epsfxsize=70mm
  \epsfbox{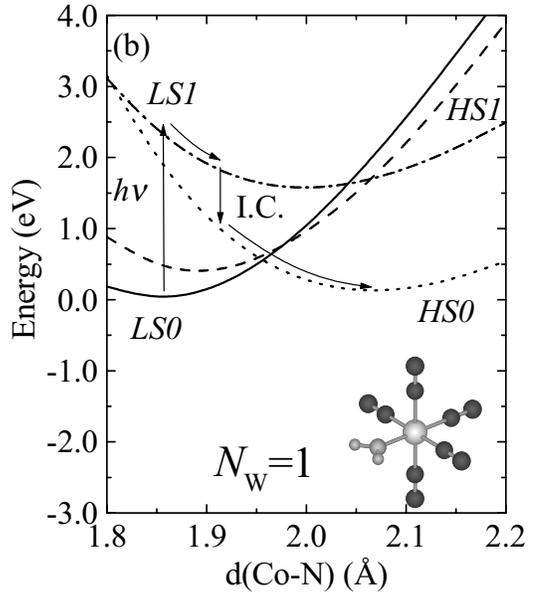}
\end{center}
\caption{
Calculated potentials of the cobalt-centered clusters with
(a) $N_{\rm W}$=0 and (b) $N_{\rm W}$=1.
The LS0, LS1, HS0, HS1 states are shown 
by solid, broken-dotted, dotted, and broken lines, respectively.
Possible primary relaxation paths for the LS0$\rightarrow$HS0 and
HS0$\rightarrow$LS0 transitions are schematically indicated by arrows
in (b) and (a), respectively (see text).
}
\label{fig:pot}
\end{figure}

Figure ~\ref{fig:pot} shows calculated cluster potentials
as functions of $d$(Co-N) for $N_{\rm W}=0$ and 1.
We see that the energies of the charge-transferred states (LS1 and HS0) are
lowered in the cluster of $N_{\rm W}=1$ compared with the case of $N_{\rm
W}=0$.
They become further stabilized in the cluster of $N_{\rm W}=2$ (not shown here).
This implies that the ligand substitution enhances the electron affinity of Co.
On the other hand, the energy differences between different spin states
(between LS0 and HS1 and between HS0 and LS1) are rather insensitive to
$N_{\rm W}$, indicating that the ligand substitution does not affect the
Hund coupling very much.
From the calculated cluster potentials we obtain the Frank-Condon excitation
energies of the CT excitations (LS0$\rightarrow$LS1 and
HS0$\rightarrow$HS1).
The results
are summarized in Table~\ref{table:absorption} together with the observed
main absorption peaks and the excitation energies responsible for the 
structural change.
The calculated excitation energy $\sim2.3$ eV of LS0$\rightarrow$LH1 in
Co($N_{\rm W}$=1)
reasonably corresponds to the observed absorption energy 2.4 eV inducing 
the magnetization. 
This implies that the photoinduced LS$\rightarrow$HS transition is
triggered by CT excitation mainly between a Co atom with $N_{\rm W}$=1 and a
nearby Fe atom.
Note that the largest fraction (37\%) of Co is in the configuration of
$N_{\rm W}$=1 
in K$_{0.4}$Co$_{1.3}$Fe(CN)$_6$, 
if iron vacancies are randomly distributed in the crystal.
The Co atoms with $N_{\rm W}$=0 (21\% fraction) 
are disadvantageous for this process because
of the large energy difference between the metastable HS0 state and the
stable LS0 state with a small potential barrier for HS0, as one can see in
Fig.~\ref{fig:pot}(a).

In the case of the reverse (HS$\rightarrow$LS) transition, the experimental
photon energy (0.9 eV) required for demagnetization is much lower than the
energy of the observed broad absorption around 2.3 eV.
The latter may correspond to the calculated CT excitations at $\sim$1.4 eV in
Co($N_{\rm W}$=1) and at $\sim$2.6 eV in Co($N_{\rm W}$=2), whereas
the former is assigned to the
excitation energy of Co($N_{\rm W}$=0) at $\sim$0.8 eV.
This implies that the HS$\rightarrow$LS transition is triggered mainly by
the excitation of Co atoms with $N_{\rm W}$=0, in contrast to the forward
transition.
This is reasonable, because the final LS0 state is much lower in energy than
the initial HS0 state in the case of $N_{\rm W}$=0 shown in
Fig.~\ref{fig:pot}(a). 

The suggested main relaxation paths in the reversible transitions are
indicated by arrows
in Fig.~\ref{fig:pot}.
The path for the LS$\rightarrow$HS transition is shown in the case of
$N_{\rm W}$=1, and the path for the reverse transition is indicated in the
case of $N_{\rm W}$=0.
The Frank-Condon photo-excitations induce CT between cobalt
and iron cations.
% modified
The intermediate excited states are subsequently transformed into local
meta-stable states via intersystem crossing due to the spin-orbit interaction
at cobalt site.
The magnitude of the spin orbit coupling has been estimated 
as about 10 meV\cite{Figgis66book}. 
This is much smaller than that of the typical energy scale of
the potential diagram in Fig. \ref{fig:pot}. 
On the other hand, it corresponds to a time scale less than 1ps, 
which is much faster than the time scale of the photoinduced phase transition.
Experimentally, intersystem crossing has been observed to occur with large
probabilities in similar cobalt complexes\cite{McCusker93IC}. 

The most important conclusion from our results is that the forward and the
backward changes are initiated from spatially different local
regions in the crystal.
The difference lies in the cluster potential for Co electrons due to the
difference in the number of ligand substitutions: $N_{\rm W}$=0 or 1.
The large concentration of Fe vacancies in this crystal, 
which might otherwise be thought to be a shortcoming of this material, 
is an essential prerequisite for the reversible transformations. 
The present model also implies that the photoinduced change 
must be a cooperative 
phenomenon\cite{Ogawa00PRL,Koshihara92PRL,Koshihara99JPCB,Yu97JMatSci}, 
because the structural change of the entire crystal is induced by 
excitations in a fraction of the crystal. 
To confirm our model, a systematic experimental study with varying $x$ 
in K$_{1-2x}$Co$_{1+x}$Fe(CN)$_6$ is desired especially.   
For instance, 
the oscillator strength of the 3.9eV absorption peak in the LS state 
will be much more enhanced for smaller $x$ because of an increased 
fraction of Co($N_{\rm W}$=0).

%::::::::::::::::::::::::::::::::::::::::::::::::::::::::::::::::::::::::::::::
%                                 discussion
%::::::::::::::::::::::::::::::::::::::::::::::::::::::::::::::::::::::::::::::

Based on the present results, we are able to speculate 
how the initial relaxation process leads to the phase transition in
the macroscopic scale. 
A local relaxed CT state (exciton) after the primary process is
still in an excited state with a finite life time in the crystal, because it
involves a strain energy due to lattice mismatch or frustration.
The strain energy is released when the surrounding lattice undergoes a
global change as an integration of all the local changes.
There should be a critical concentration, $R_{\rm C}$, of relaxed
excitons, above which
the fraction of the new phase increases drastically in the crystal. 
Such behavior was observed in a similar photo-induced phase transition of a 
spin-crossover complex\cite{Ogawa00PRL}.
The exciton concentration $R$ satisfies
$P^{\rm PI } ( 1 - R) = R/{\tau}$
in the stationary condition,
where $P^{\rm PI}$ and $\tau$ represents the creation rate
and the lifetime of exciton, respectively.
The critical excitation rate $P^{\rm PI }_{\rm C}$ is then expressed as
$P^{\rm PI }_{\rm C} = R_{\rm C}/[\tau(1-R_{\rm C})]$.
Therefore, the phase transition occurs more easily 
with a smaller excitation light
intensity for a longer exciton lifetime.
In the present system, the long life time of the relaxed exciton is
guaranteed by its local stability as shown in Fig.~\ref{fig:pot}. 

The local mechanism discussed here indicates a general strategy to realize a
reversible photoinduced phase transition by the use of a mixed crystal of
$a$ and $b$ molecules.
(Here a `molecule' means just a structural unit in a wide sense.)
Suppose both the $a$ and $b$ molecules are bistable between the two types of state A
and B, but they are different in that the $a$ molecule prefers the A-type
state and the $b$ molecule prefers the B-type state.
(In the Prussian blue analogues the clusters with $N_{\rm W}$=1 and 0 
play the roles of the $a$ and $b$ molecules.) 
In the A phase of the crystal where all the molecules are in the A-type
states, the $b$ molecules are metastable locally.
By selectively exciting $b$ molecules by light, their states are readily
changed to the locally stable B-type states, triggering the phase transition
of the entire crystal to the B phase where all the molecules are in the
B-type states.
In the B phase, the story is the same with the interchanged roles of the $a$
and the $b$ molecules: Photo-excitation of locally metastable $a$ molecules
triggers the phase transition to the A phase.

In summary, our ab initio calculations have elucidated that 
the photoinduced forward and backward phase transitions are 
initiated by excitation of different local regions by different photons
in Prussian blue analogues.
Especially, the existence of a large concentration of vacancies, 
which might be thought as a shortcoming of this material, 
turned out to be essential for the bidirectional phase transition.  
We have also proposed that this mechanism can be generalized 
as an effective strategy to design a new material for bidirectional 
photoinduced switching. 

The authors acknowledge Prof. K. Hashimoto at Tokyo University and
Dr. O. Sato at Kanagawa Academy of Science and Technology for useful
discussions. One of the authors (Y.A.) appreciates
Prof. T. Iyoda at Tokyo Metropolitan University for communications in the
initial stage of our study.
Computations have been done partly using the
facilities of AIST Tsukuba Advanced Computing Center.

%\bibliographystyle{prsty}
%\bibliography{prussian}

\begin{thebibliography}{10}

\bibitem{Sato97JES}
O. Sato, Y. Einaga, T. Iyoda, A. Fujishima, and K. Hashimoto, J. Electrochem.
  Soc. {\bf 144},  L11  (1997).

\bibitem{Sato96Science}
O. Sato, T. Iyoda, A. Fujishima, and K. Hashimoto, Science {\bf 272},  704
  (1996).

\bibitem{Verdaguer96Science}
M. Verdaguer, Science {\bf 272},  698  (1996).

\bibitem{Goujon00EPJB}
A. Goujon, O. Roubeau, F. Varret, A. Dolbecq, A. Bleuzen, and M. Verdaguer,
  Eur. Phys. J. B {\bf 14},  115  (2000).

\bibitem{Pejakovic00PRL}
D.~A. Pejakovi\'{c}, J.~L. Manson, J.~S. Miller, and A.~J. Epstein, Phys. Rev.
  Lett. {\bf 85},  1994  (2000).

\bibitem{Yoshizawa98JPCB}
K. Yoshizawa, F. Mohri, G. Nuspl, and T. Yamabe, J. Phys. Chem. B {\bf 102},
  5432  (1998).

\bibitem{Kawamoto99PRB}
T. Kawamoto, Y. Asai, and S. Abe, Phys. Rev. B {\bf 60},  12990  (1999).

\bibitem{Nishino98PRB}
M. Nishino, K. Yamaguchi, and S. Miyashita, Phys. Rev. B {\bf 58},  9303
  (1998).

\bibitem{Ferlay95Nature}
S. Ferlay, T. Mallah, R. Ouah\`{e}s, P. Veillet, and M. Verdaguer, Nature {\bf
  378},  701  (1995).

\bibitem{Sato96Science2}
O. Sato, T. Iyoda, A. Fujishima, and K. Hashimoto, Science {\bf 271},  49
  (1996).

\bibitem{Ohkoshi98JACS}
S. Ohkoshi, A. Fujishima, and K. Hashimoto, J. Am. Chem. Soc. {\bf 120},  5349
  .

\bibitem{Ohkoshi99JACS}
S. Ohkoshi and K. Hashimoto, J. Am. Chem. Soc. {\bf 121},  10591  (1999).

\bibitem{Yokoyama98PRB}
T. Yokoyama, T. Ohta, O. Sato, and K. Hashimoto, Phys. Rev. B {\bf 58},  8257
  (1998).

\bibitem{Decurtins84CPL}
S. Decurtins, P. G\"{u}tlich, C.P. Kohler, H. Spiering, and A. Hauser, Chem. Phys.
  Lett. {\bf 105},  1  (1984).

\bibitem{Ogawa00PRL}
Y. Ogawa, S. Koshihara, Koshino. K., T. Ogawa, C. Urano, and H. Takagi, Phys.
  Rev. Lett. {\bf 84},  3181  (2000).

\bibitem{Schmidt93JCompPhys}
M.W. Schmidt, K.K. Baldridge, J.A. Boatz, S.T. Elbert, M.S. Gordon, J.H.
  Jensen, S.Matsunaga Koseki, N. Nguyen, K.A. Su, S. Windus, T.L. Dupuis, M.
  Montgomery, and Jr. J.A., J. Comput. Chem. {\bf 14},  1347  (1993).

\bibitem{Huzinaga84book}
{\em Gaussian basis set for molecular calculations}, edited by S. Huzinaga
  (Elsevier, Amsterdam, 1984).

\bibitem{Juszczyk94JPhys}
S. Juszczyk, C. Johansson, M. Hanson, A. Ratuszna, and G. Malecki, J. Phys.,
  Condens. Matter. {\bf 6},  5697  (1994).

\bibitem{Figgis66book}
{\em Introduction to LIGAND FIELDS}, edited by B.~N. Figgis (Interscience
  Publishers, New York, London, Sydney, 1966).

\bibitem{McCusker93IC}
J.~K. McCusker, K.~N. Walda, D. Magde, and N. Hendrickson, Inorg. Chem. {\bf
  32},  394  (1993).

\bibitem{Koshihara92PRL}
S. Koshihara, Y. Tokura, K. Takeda, and T. Koda, Phys. Rev. Lett. {\bf 68},
  1148  (1992).

\bibitem{Koshihara99JPCB}
S. Koshihara, Y. Takahashi, H. Sakai, Y. Tokura, and T. Luty, J. Phys. Chem. B
  {\bf 103},  2592  (1999).

\bibitem{Yu97JMatSci}
Z. Yu, Y.F. Hsia, X.Z. You, H. Spiering, and P. G\"{u}tlich, J. Mater. Sci. {\bf
  32},  6579  (1997).

\end{thebibliography}

\begin{table}
\caption{
Frank-Condon excitation energies of the LS and HS states evaluated from the
cluster potentials (as shown in Fig. 3) for various $N_{\rm W}$,
compared with experiments of
the absorption peak energies (underlined) and the excitation energies of the
phase transitions (in bold).
}
\label{table:absorption}
\begin{center}
\begin{minipage}{8.5cm}
\begin{tabular}{llcc}
                   & & \multicolumn{2}{c}{Excitation Energy(eV)} \\ \hline
 Theory                   & &   LS0$\rightarrow$LS1  & HS0$\rightarrow$HS1  \\ 
\multicolumn{1}{r}{$N_{\rm W}$=0}& &  3.9               & {\bf 0.8} \\
  \multicolumn{1}{r}{=1}& &  \underline{{\bf 2.3}}  & \underline{1.4} \\ 
  \multicolumn{1}{r}{=2}& &    0.7                  & \underline{2.6} \\\hline
 Experiment             &&                       &                 \\
 \hspace{5mm}Absorption peak        &&  \underline{2.4}   & \underline{2.3} \\
 \hspace{5mm}Photo-induced change   &&  {\bf 2.4}         & {\bf 0.9} \\
\end{tabular}
\end{minipage}
\end{center}
\end{table}

\clearpage

\end{document}